\documentstyle[dina4,epsf,twocolumn]{article}
\begin{document}
\renewcommand{\textfraction}{0.0}
\renewcommand{\topfraction}{1.0}
\renewcommand{\bottomfraction}{1.0}
\def\figurename{Fig.}
\newcommand{\gapro}
  {\raisebox{-0.25ex} {$\,\stackrel{\scriptscriptstyle>}%
    {\scriptscriptstyle\sim}\,$}}
\newcommand{\lapro}
   {\raisebox{-0.25ex} {$\,\stackrel{\scriptscriptstyle<}%
    {\scriptscriptstyle\sim}\,$}}
\rule[-8mm]{0mm}{8mm}
\begin{minipage}[t]{16cm}
{\large \bf ON THE STABILITY OF POLARONIC SUPERLATTICES IN 
\\ STRONGLY COUPLED ELECTRON--PHONON SYSTEMS\\[4mm]}
G. Wellein$^1$,  H.~Fehske$^1$, H. B\"uttner$^1$, 
and A. R. Bishop$^2$\\[3mm]
$^1$Physikalisches Institut, Universit\"at Bayreuth, D--95440 Bayreuth,
Germany\\
$^2$MS B262 Los Alamos National
Laboratory, Los Alamos, NM 87545, USA\\[4.5mm]
\hspace*{0.5cm}
We investigate the interplay of electron--phonon (EP) coupling 
and strong electronic correlations in the frame of 
the two--dimensional (2D) Holstein t--J model (HtJM), 
focusing on polaronic ordering 
phenomena for the quarter--filled band case.  
The use of direct Lanczos diagonalization on finite lattices 
allows us to include the effects of quantum phonon fluctuations 
in the calculation of spin/charge structure factors and hole--phonon 
correlation functions.  
In the adiabatic strong coupling regime we found evidence for  
``self-localization'' of polaronic carriers in a $(\pi,\pi)$ charge-modulated
structure, a type of superlattice solidification reminiscent of those 
observed in the nickel perovskites $\rm La_{2-x}Sr_{x}NiO_{4+y}$.
\end{minipage}\\[4.5mm]
\normalsize
Electron diffraction measurements on the nickel oxide analogue 
of the layered high--$T_c$ 214 copper oxide compound,   
$\rm La_{2-x}Ni_xCuO_4$, reveal a $(\pi,\pi)$--superstructure 
spot at {\it quarter filling}, i.e. for $x=0.5$, 
which has been interpreted as sign of a truly 
2D polaron ordered phase~[1]. 

To study charge-- and spin--ordering phenomena 
in such systems, exhibiting besides a substantial EP interaction 
strong Coulomb correlations, let us consider the planar t-J model 
with an additional on-site Holstein hole-phonon coupling:
\begin{eqnarray}
{\cal H}^{}_{H-t-J}\!\!&=&\!\!{\cal H}_{ph}^{}+{\cal H}^{}_{ho-ph}
+{\cal H}^{}_{t-J}\,,\\
{\cal H}^{}_{ho-ph}\!\!&=&\!\!- \sqrt{\varepsilon_p\hbar\omega_0}  
\sum_i \big(b_i^\dagger + b_i^{}
\big)\,\tilde{h}_i^{}\,.
\end{eqnarray} 
Here ${\cal H}_{ph}$ and ${\cal H}^{}_{t-J}$ represent the phonon part
and standard t--J model, respectively. 
In~(2), $\varepsilon_p$ is the local EP coupling constant, 
$\omega_0$ denotes the bare phonon frequency,  $\tilde{h}_i=1-\sum_\sigma
\tilde{c}_{i\sigma}^\dagger \tilde{c}_{i\sigma}^{}$, and $b_i^\dagger$ 
($\tilde{c}_{i\sigma}^\dagger$) are the usual phonon (projected fermion)
creation operators. 
Of course the HtJM gives only a 
very simplified description of the complex hole 
transport and low--spin high--spin 
interactions in the nickelates~[2].     
\begin{figure}
\rule[-5.7cm]{0mm}{6.8cm}
\end{figure}

At quarter--filling and for strong EP interactions, 
Lanczos studies of the {\it static} HtJM 
(frozen--phonon approximation) yield strong indications 
of a Peierls distorted ground state~[3] . 

To discuss {\it non-adiabatic} effects preserving 
the full dynamics and quantum nature of the phonon degrees of freedom, 
we perform here a direct exact diagonalization (ED) of the HtJM 
on a ten-site ($N$) square lattice with at most $M$~phonons 
using a well--controlled phononic Hilbert space truncation procedure~[4].
Since memory limitations impose severe restrictions 
on this method, we study an effective polaronic t--J model
${\cal H}_{H-t-J}^{eff}({\mit \Delta}_i, \gamma,
\bar{\gamma}, \tau^2)$ as well, which can be derived from~(1) by applying 
the \mbox{inhomogeneous} modified variational Lang--Firsov (IMVLF) approach 
outlined in~[4]. The $\!N\!+\!2$ \mbox{variational} parameters
take into \mbox{account} {\it static} 
displacement field (${\mit \Delta}_i$), {\it dynamic} polaron ($\gamma$), 
finite density ($\bar{\gamma}$) and squeezing $(\!\tau^2)$~effects. 
We stress that the IMVLF-Lanczos approach correctly reproduces
the adiabatic and anti-adiabatic, weak- and strong EP coupling 
limits$\!$~[5]. 

In the numerical analysis of the HtJM, we first consider the 
case of spinless fermions (total $S^z=S^z_{max}$), i.e.,  
the electronic correlations are neglected. Increasing the EP coupling 
at fixed phonon frequency  ($\hbar\omega_0=0.8$), the smooth variation of the
charge structure factor $S_c(\pi,\pi)$ in the nearly free polaron  
state ($\varepsilon_p\lapro 2$) is followed by a strong enhancement in 
a ``quasi--localized'' polaron state indicating the formation 
of a charge density wave (CDW) [see Fig.~1]. 
This crossover becomes suppressed in the non--adiabatic regime 
($\hbar\omega_0=3$). As can be seen from the insets, at $\varepsilon_p=3$, i.e.
in the CDW--like phase, a larger number of 
phonons is still required to achieve a satisfactory 
convergence of the ED data. 
Including more phonons we expect an even more pronounced increase 
of $S_c(\pi,\pi)$ [cf.~the~IMVLF~curve].
\begin{figure}[t]
\centerline{\mbox{\epsfxsize 8cm\epsffile{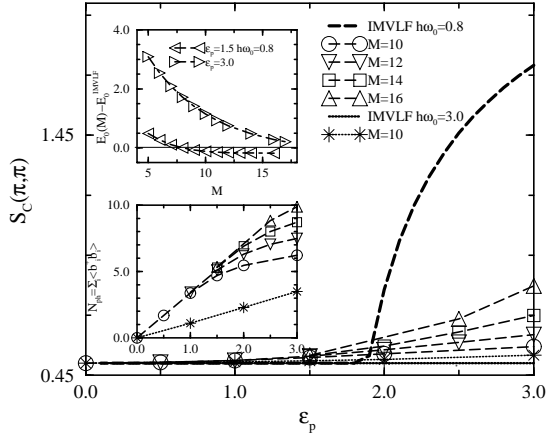}}}
\vspace*{-0.3cm}
\caption{Charge structure factor $S_c(\vec{q})$ vs $\varepsilon_p$, where 
ED data with different $M$ are compared with 
IMVLF results. The insets show the deviation of $E_0(M)$ from
the IMVLF--value and the mean number of phonons $N_{ph}$ in the ground state, 
respectively. All energies are measured in units of $t$ 
and we adopt $J=0.4$ throughout.}\vspace*{-0.3cm}
\end{figure}
\begin{figure}[b]
\vspace*{-0.3cm}
\centerline{\mbox{\epsfxsize 8cm\epsffile{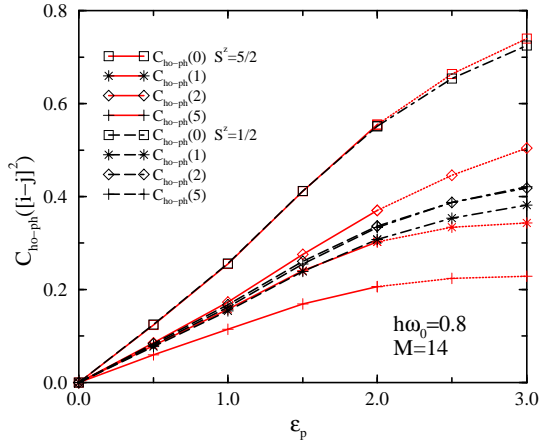}}}
\vspace*{-0.3cm}
\caption{$C_{ho-ph}(|i-j|)$ vs $\varepsilon_p$ 
at all nonequivalent distances 
$|\protect\vec{R}_i-\protect\vec{R}_j|/a=0, 1, \protect\sqrt{2},
\protect\sqrt{5}$. Increasing $M$ for 
$\protect\varepsilon_p \protect\gapro 2$, we expect an upturn 
of $C_{ho-ph}(0,2)$.\protect\vspace{-0.1cm}}
\end{figure}

To visualize the hole--phonon correlations, Fig.~2
displays the variation of $C_{ho-ph}(|i-j|)=\langle 
{\mit \Psi}_0|\tilde{h}_ib_j^\dagger b_j^{}|
{\mit \Psi}_0\rangle$ as a function of $\varepsilon_p$ for both the 
spin-1/2 and spinless fermion cases. Our results clearly 
show the phonon dressing of the holes according to an AB  
sublattice structure. Due to the possible gain of 
exchange energy~$J$, the density oscillations are weakened for 
spin-1/2~particles.

Using the IMVLF-Lanczos method to reach the strong--coupling
adiabatic regime, we notice, as 
$\varepsilon_p$ increases, a sequence of transitions 
from nearly free to self--trapped polarons 
[with less mobility $\propto t_{p,eff}=
E_{\rm kin}(\varepsilon_p,J)/E_{\rm kin}(0,J)$]
solidifying into a polaronic superlattice and finally 
to charge--separated (CS) states ($\varepsilon_p\gapro 3$).
These transitions are accompanied by a  
change of both charge {\it and} spin structure factors [see Fig.~3]. 
Note that the spin correlations are significantly 
enhanced (weakened) in the CS (CDW) state.
\begin{figure}[t]
\centerline{\mbox{\epsfxsize 8cm\epsffile{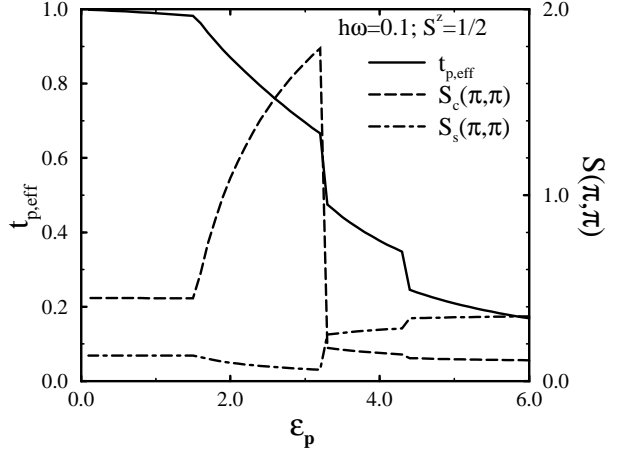}}}
\vspace*{-0.3cm}
\caption{Magnetic/charge structure factors $S_{c/s}(\vec{q})$ and $t_{p,eff}$
vs $\varepsilon_p$ for the 2D HtJM at quarter filling. }\vspace*{-0.3cm}
\end{figure}

This work was supported by DFG, SFB 279.
G.$\!\!$~W. and H.$\!$~F. would like to thank the$\!$~Theoretical Division 
of LANL for hospitality. Work at LANL is performed under the auspices
of the US DOE. 
\\[0.17cm]
REFERENCES
\begin{enumerate}\vspace*{-0.1cm} 
\item S.$\!$ Cheong$\!$ {\it et al.}, Phys.$\!$ Rev.$\!$ B$\!$ {\bf 49},
$\!$ 7088$\!$ (1994);
 J.$\!$ Tranquada$\!$  {\it et al.}, Nature$\!$ {\bf 375}, 561 (1995).
\vspace*{-0.18cm}
\item J. Loos and H. Fehske, Czech. J. Phys. {\bf 46}, 1879 (1996).
\vspace*{-0.18cm}
\item J. Zhong and H.~Sch\"uttler, Phys. Rev. Lett. {\bf 69}, 1600
(1992); H. R\"oder, H. Fehske, and R. N. Silver, 
Europhys. Lett. {\bf 28}, 257 (1994).
\vspace*{-0.18cm}
\item G. Wellein, H. R\"oder, and H. Fehske, Phys. Rev. B {\bf 53}, 
9666 (1996).
\vspace*{-0.18cm} 
\item H.$\!$ Fehske$\!$ {\it et al.},
Phys.$\!$ Rev.$\!$ B$\!$ {\bf 51},$\!$ 16582$\!$ (1995).
\end{enumerate}
\end{document}